\documentclass[10pt,letterpaper]{article}
\usepackage[top=0.85in,left=2.75in,footskip=0.75in]{geometry}

\usepackage{amsmath,amssymb}

\usepackage{changepage}

\usepackage{textcomp,marvosym}

\usepackage{cite}

\usepackage{nameref,hyperref}

\usepackage[right]{lineno}

\usepackage[nopatch=eqnum]{microtype}
\DisableLigatures[f]{encoding = *, family = * }

\usepackage[table]{xcolor}

\usepackage{array}

\newcolumntype{+}{!{\vrule width 2pt}}

\newlength\savedwidth

\raggedright
\usepackage[aboveskip=1pt,labelfont=bf,labelsep=period,justification=raggedright,singlelinecheck=off]{caption}

\makeatletter
\renewcommand{\@biblabel}[1]{\quad#1.}
\makeatother

\usepackage{lastpage,fancyhdr,graphicx}
\usepackage{epstopdf}
\fancyheadoffset[L]{2.25in}
\fancyfootoffset[L]{2.25in}
\begin{document}
\vspace*{0.2in}

\begin{flushleft}
{\Large
\textbf\newline{Benchmarking spike source localization algorithms in high density probes}
}
\newline
\\
Hao Zhao*\textsuperscript{1},
Xinhe Zhang*\textsuperscript{1,2},
Arnau Marin-Llobet\textsuperscript{1},
Xinyi Lin\textsuperscript{1}, and
Jia Liu\textsuperscript{1}
\\
\bigskip
\textbf{1} John A. Paulson School of Engineering and Applied Sciences, Harvard University, Boston, MA, USA \\
\textbf{2} Broad Institute of MIT and Harvard, Cambridge, MA, USA \\
\bigskip

* Corresponding authors: hzhao@mba2024.hbs.edu, xinhezhang@g.harvard.edu

\end{flushleft}

\section*{Abstract}
Estimating neuron location from extracellular recordings is essential for developing advanced brain-machine interfaces. Accurate neuron localization improves spike sorting, which involves detecting action potentials and assigning them to individual neurons. It also helps monitor probe drift, which affects long-term probe reliability. Although several localization algorithms are currently in use, the field is nascent and arguments for using one algorithm over another are largely theoretical or based on visual analysis of clustering results. We present a first-of-its-kind benchmarking of commonly used neuron localization algorithms. We tested these algorithms using two ground truth datasets: a biophysically realistic simulated dataset, and experimental data combining patch-clamp and Neuropixels probes. We systematically evaluate the accuracy, robustness, and runtime of these algorithms in ideal conditions and long-term recording conditions with electrode decay. Our findings highlight significant performance differences; while more complex and physically realistic models perform better in ideal situations, simple heuristics demonstrate superior robustness to noise and electrode degradation in experimental datasets, making them more suitable for long-term neural recordings. This work provides a framework for assessing localization algorithms and developing robust, biologically grounded algorithms to advance the development of brain-machine interfaces.

\section*{Author summary}
Accurately estimating neuron locations from extracellular recordings is critical to building reliable brain–machine interfaces. This spatial information enhances spike sorting and enables long-term monitoring of neural activity, especially in the presence of probe drift and electrode degradation. Despite the availability of several spike source localization algorithms, their comparative long-term performance has not been systematically benchmarked against ground truth data. In this study, we benchmark three widely used algorithms—center of mass (COM), monopolar triangulation (MT), and grid convolution (GC)—using both simulated and experimental datasets. We assess their accuracy, runtime, and robustness under ideal and degraded recording conditions. Our results reveal that while MT and GC offer higher accuracy in ideal conditions, COM demonstrates superior resilience to noise and electrode decay, making it more suitable for long-term recordings. These findings provide a foundational framework for evaluating and improving spike localization algorithms and highlight the importance of robustness in real-world neural interface applications.

\section*{Introduction}
Improvements in high-density multi-electrodes arrays (MEAs) in the past several years have enabled the extracellular study of neuron populations over long periods of time \cite{van Beest 2024, Zhao 2023}. A key advantage of high-density MEAs is the spatial density of electrodes, which allows for the same action potential (or “spike”) to be detected at multiple electrode sites, providing an opportunity to estimate the location of each spike. This spatial information is important in a variety of downstream tasks. One of these key tasks is “spike sorting”, which is the process of detecting these action potentials and assigning each spike to an individual neuron or “unit”. Accurate spike sorting is crucial for understanding neural circuits, as incorrectly assigned spikes can significantly distort the interpretation of neural activity \cite{Quiroga 2012}. Currently, state-of-the-art spike sorting algorithms and strategies rely heavily on spatial information to enhance their accuracy \cite{Pachitariu 2023, He 2024, Garcia 2024, Lin 2025B}. Following the spike sorting process, the estimated location of the neuron (based on the aggregate signal of all spikes assigned to that neuron, referred to as a “template”) is useful for the evaluation of brain-machine interface performance (e.g., to assess probe drift) \cite{Boussard 2021}.

While several localization algorithms are currently used, there has been no ground-truth benchmarking of their performance and robustness against decay. The most widely used localization algorithm is center of mass (“COM”), which uses a “weighted average” heuristic that is simple but physically inaccurate. Other more physically accurate algorithms have been proposed in recent years, including monopolar triangulation (“MT”) and grid convolution (“GC”); along with COM, these methods are implemented in the widely used SpikeInterface package \cite{Garcia 2022, Boussard 2021}. However, despite significant variance in the algorithms and their underlying physical assumptions, there has been no systematic analysis to date of their performance in long-term recording situations. 

A better understanding of performance is particularly important given advancements in neural probe technology. Current probes, such as those which are designed with tissue-level flexibility, are able to record from the same neuron populations over periods of up to over a year \cite{Tang 2023, Araki 2020, Sharafkhani 2022, Xu 2022, Yasar 2024, Sheng 2025, Liu 2024}. Over long-term recording horizons, electrodes within the array will decay, and localization algorithms should be robust against this loss of signal if they are to be suitable for decoding long-term brain recordings and accurately track the same neurons over long periods of time. Additionally, probes with tissue-level flexibility reduce the mechanical and structural differences between the probe and brain tissue, and appropriate localization algorithms are important to assess the degree of improvement in probe drift \cite{Tang 2023, Lee 2023, Lin 2025, Zhao 2023}.

To address this lack of understanding, we perform a first-of-its-kind benchmarking of COM, MT, and GC against two sources of ground truth data and electrode decay. The first is a biophysically realistic simulated dataset, which has recently been used to benchmark stabilization algorithms for probe drift (a separate but related problem). The second is an experimental dataset of paired recordings where the same neurons were recorded from patch clamp (providing the “ground truth” spike train using intracellular data) and multi-electrode array (Neuropixel 384-channel probe), and the approximate neuron location is known for each paired recording. For each dataset, we first benchmark the performance of all three localization algorithms on both spikes and templates. Then, we simulate electrode decay and examine how robust the different localization algorithms are to decay.

\section*{Materials and methods}
\subsection*{Simulated Dataset}

Our simulated dataset uses the MEArec simulator to generate ground truth recordings using biophysically detailed multicompartment models, and follows similar protocols to previous analyses used to simulate ground truth conditions \cite{Garcia 2024, Buccino 2021}. We use a dictionary of 13 cell models from layer 5 of a juvenile rat somatosensory cortex to simulate a dictionary of biologically plausible templates on a 25-probe layout (5 rows and 5 columns, spaced 30nm apart). For each recording, we simulate 10 neurons selected from our template dictionary and generate corresponding spike trains. Templates and spike trains are then convolved, adding a slight modulation in amplitude to add physiological variability. Finally, uncorrelated Gaussian noise with 1 mV standard deviation was added to the traces. The sampling rate of the simulated recordings is 30 kHz.

From the simulated dataset, we know the ground-truth locations and spike trains of each neuron, and we use these ground-truth spike trains to extract spike waveforms and template waveforms (aggregations of the spike waveforms for a given neuron) at each recording electrode.

\subsection*{Experimental Dataset}
For experimental benchmarking, we use the open-source SPE-1 dataset of paired patch clamp and MEA (Neuropixel 384-channel) recordings from the same cortical neurons of anaesthetized rats \cite{MarquesSmith 2020}. The patch clamp recordings are used to extract the ground-truth spike train for the neuron, and we then use this ground-truth spike train to extract spike waveforms and template waveforms at each recording electrode. In this dataset, the authors additionally tracked the relative location of the patch-clamp to the MEA in each paired recording, and were able to approximate the closest electrode (on the MEA) to the patch clamp. The dataset consists of 43 paired recordings, of which we select 11 pairs with the strongest signal quality and confidence of patch-clamp and MEA pairing.

\newgeometry{margin=1cm}
\begin{figure}[ht!]
    \centering
    \includegraphics[width=\dimexpr\linewidth\relax, clip]{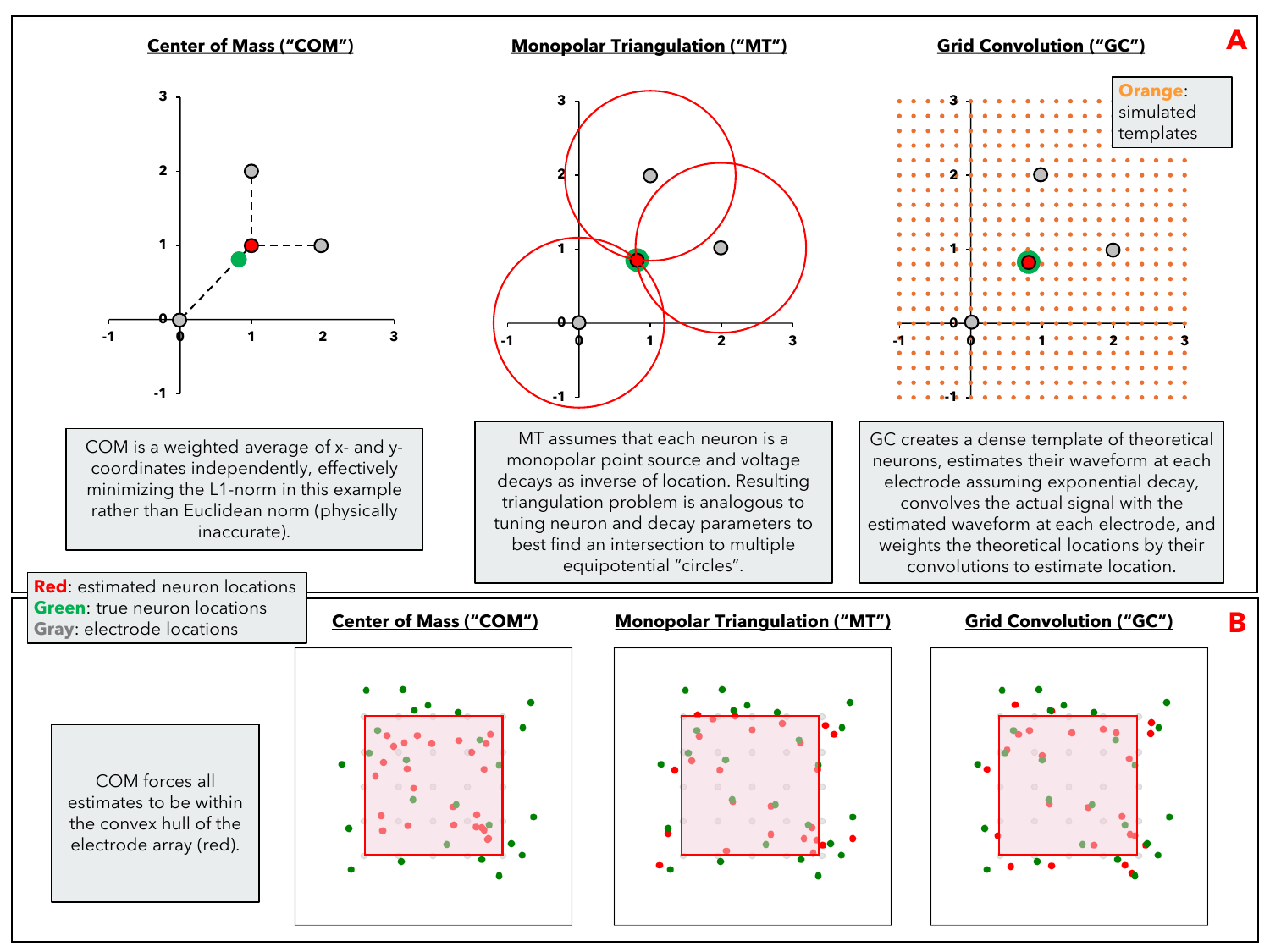}
    \caption{Limitations of center of mass localization. \textbf{A}: Visualization of localization algorithms for a point-source neuron equidistant to three electrodes located on a grid at (0,0), (1,2), and (2,1), where the electrodes will all receive the same signal strength from the neuron in ideal recording conditions (no noise or electrode decay), given the neuron is located equidistant to all three electrodes. MT and GC are able to correctly estimate the “true” location while COM is not. \textbf{B}: Simulated electrophysiological data (MEArec), where green dots denote true neuron locations. Since COM is a weighted average of electrode locations, it forces all location estimates to be within the convex hull of the electrode array (all red dots, denoting estimated neuron locations, are within red square).}
    \label{fig:fig1}
\end{figure}
\restoregeometry

\subsection*{Localization algorithms}
For spike \(i\), we denote the spike waveform at each electrode \(j\) as \(w_{ij}(t)\). We define the peak-to-peak amplitude of \(w_{ij}(t)\) as the voltage difference between the peak and the trough of the waveform, or \(\mathrm{ptp}_{ij}\). For brevity, we denote the spike location as \(p_i = (x_i, y_i, z_i)\) and electrode location as \(p_j = (x_j, y_j, z_j)\) (the subscript will indicate whether we are referring to a spike or electrode location). We will define \(x\) and \(y\) to lie along the  two-dimensional plane of the MEA, and \(z\) to lie on the vector orthogonal to the MEA. See Fig. 1 for a visualization of these three localization algorithms.

\bigskip
\textbf{Center of Mass (``COM'')}: The location of spike \(i\) is estimated as:
\[
p_i = (x_i, y_i, z_i) = \frac{\sum_{j} \mathrm{ptp}_{ij} \cdot p_j}{\sum_{j} \mathrm{ptp}_{ij}}
\]

We note that COM has known theoretical limitations to estimating neuron location correctly, but it continues to be the most commonly used heuristic for spike source localization due to speed and simplicity.

\bigskip
\textbf{Monopolar Triangulation (``MT'')}: Assumes that the neuron is a monopolar point source and voltage signal decays as an inverse of the distance from the neuron \cite{Chelaru 2005, Boussard 2021, Mechler 2011}. For spike \(i\), the peak-to-peak voltage measured at each electrode can therefore be expressed as:
\[
V_{ij} \simeq \frac{c_i}{\|p_i - p_j\|_2}
\]

where \(c_i\) is a unique constant and \({\|\cdot\|_2}\) is the Euclidean norm. For a multi-electrode array of \(N\) electrodes, spike \(i\) will have \(N\) equations, representing \(V_{ij}\) at \(N\) electrodes. We define a loss function based on the difference between the estimated \(V_{ij}\) and actual \(\mathrm{ptp}_{ij}\):
\[
L_i = \sum_{j} ({\mathrm{ptp}_{ij} - V_{ij}})
\]

We then estimate the location of spike \(i\) as that which minimizes the loss function (usually via least squares).

\bigskip
\textbf{Grid Convolution (``GC'')}: Creates a dense grid of \(k\) theoretical templates of waveform \(w_k(t)\) located at positions \(p_{k}\) (the grid is usually denser than the electrodes). For each of these theoretical templates, we simulate the signal that the template would produce on each electrode of the multi-electrode array, assuming an exponential decay:
\[
\tau_{kj}(t) = e^{({p_{k}-p_{i})}^2/(2\sigma^2)}w_k(t)
\]

The dot product between \(w_{ij}(t)\) and \(\tau_{kj}(t)\) measures similarity; we can estimate the location of spike \(i\) as a weighted product of the theoretical template locations using this similarity measure \cite{Pachitariu 2023}.

\newgeometry{margin=1cm}
\begin{figure}[ht!]
    \centering
    \includegraphics[width=\dimexpr\linewidth\relax, clip]{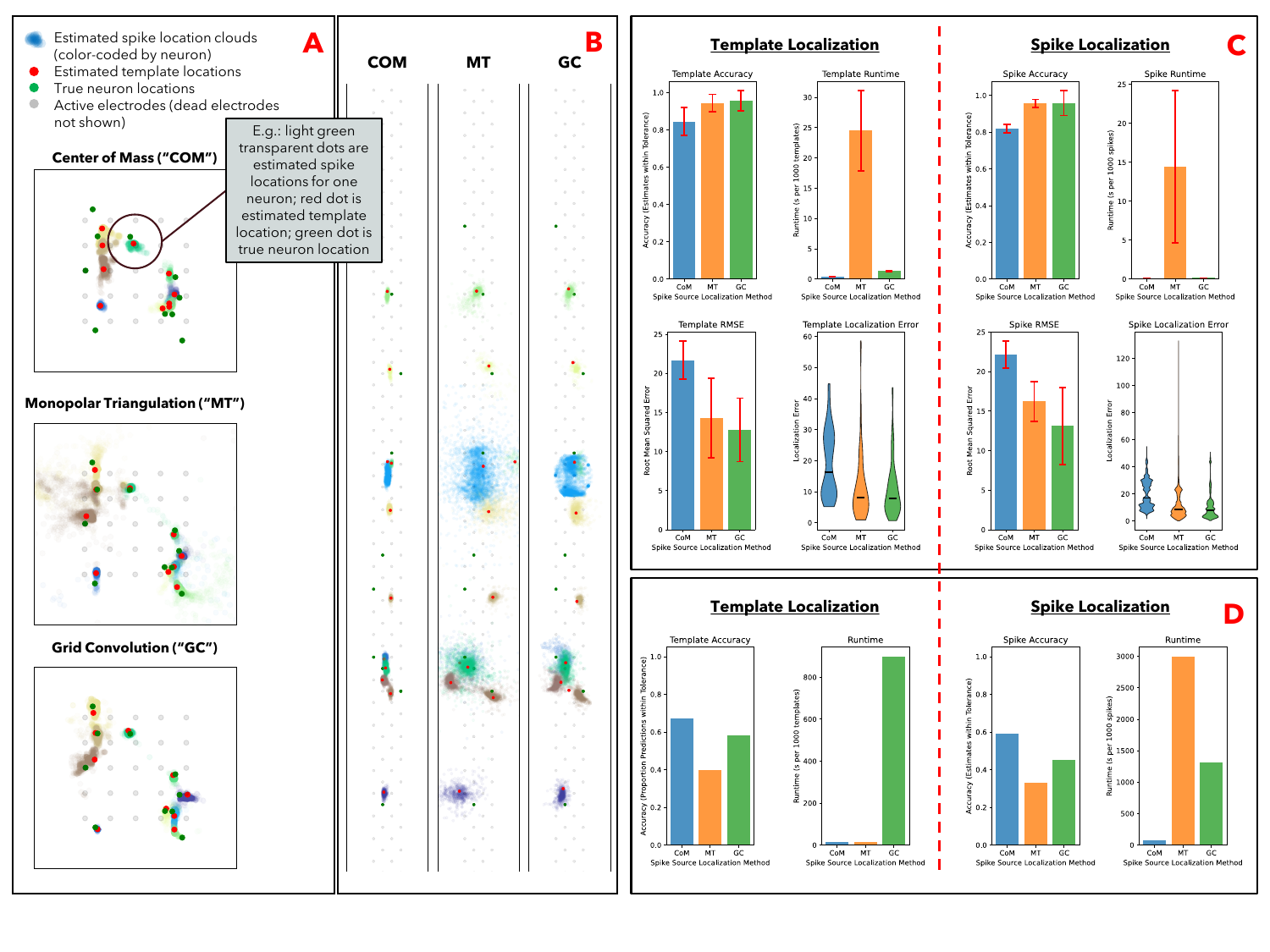}
    \caption{SSL algorithm performance with no electrode degradation. \textbf{A}: Visualization of template and spike localization estimates on simulated dataset (MEArec). The estimated spike locations are represented by translucent dot clouds, color-coded by their corresponding neuron. The estimated template location and true neuron locations are represented by solid red and green dots, respectively, and are not color-coded by neuron. The electrode array (shown excluding the “dead” electrodes representing electrode decay) is denoted by light gray dots. \textbf{B}: Visualization of template and spike localization estimates on experimental dataset (SPE-1). \textbf{C}: Performance metrics on simulated dataset, for both templates (left) and spikes (right). Performance is assessed as (i) percentage of estimates within 30µm of actual locations (accuracy), (ii) speed of algorithm on all templates/spikes (runtime), and (iii) Euclidean distance between estimates and actual location (RMSE and localization error). Bars represent mean metric across multiple simulations; error bars represent standard deviation of metric. Statistical significance was assessed using one-way ANOVA followed by Tukey’s HSD post-hoc tests; all metrics exhibited significant differences (p \textless{} 0.05). \textbf{D}. Performance metrics on experimental dataset, for both template (left) and spikes (right). Performance is assessed using accuracy and runtime; since we have lower resolution actual location in experimental data, we use a higher tolerance for accuracy (deemed correct if estimate within 80µm of actual location) and do not calculate the localization error. Bars represent metric across all experimental data.}
    \label{fig:fig2}
\end{figure}
\restoregeometry

\section*{Results}

\subsection*{Benchmarking results in absence of electrode decay}

We benchmarked all three localization algorithms using the simulated ground truth dataset (Fig. 2A) and experimental ground truth dataset (Fig. 2B) without electrode decay. In the visualization of these localization results, we see hallmarks of each algorithm consistent with prior visual analysis. We see that COM is only able to produce location estimates within the convex hull of the electrode array, and tends to cluster the estimates in the middle of the array, whereas MT and GC are able to produce more expressive estimates that appear closer to the actual neuron locations. Since we have access to the ground truth data, we are able to quantitatively benchmark the results beyond visual inspection. The metrics we benchmark for each algorithm are accuracy (percentage of estimated locations within a specified radius of the actual location), localization error (Euclidean distance between the estimated location and actual location), and runtime. In the experimental dataset, since the ground truth actual location is known with lower resolution (there is a margin of error in the instrument calibration used to obtain the data), we allow a larger radius for calculating “accurate” predictions and do not calculate localization error (where the calibration error may dominate algorithm error).

In the simulated dataset (Fig. 2C), GC and MT perform better than COM. This is consistent with our visual analysis, as COM’s tendency to cluster estimates toward the center of the array, as well as its inability to express estimates outside of the convex hull of the array, result in lower-quality estimates. In the experimental dataset (Fig. 2D), COM and GC perform better than MT. Theoretically, COM should perform relatively worse than GC and MT on the experimental probe (Neuropixel 384-channel), which is 5mm in length and 40~{\textmu}m in width, as its geometry is closer to a “linear” one-dimensional probe. In a one-dimensional probe, the convex hull of the array narrows to a line and COM is only able to estimate locations along this line, whereas the more physically realistic qualities of GC and MT leverage the relative signal strength along the line to infer locations in the two dimensions orthogonal to the simplified probe. Our belief as to why this does not play out in practice is that the experimental dataset is inherently noisier, and COM is more robust to noise than the other algorithms (as we will see in electrode decay). Lastly, we note COM is fastest in both the simulated and experimental datasets, and MT is substantially slower as it must solve an optimization problem. Speed is important in spike localization (and less so for templates) given the higher volume of spikes and the need to localize quickly for real-time spike sorting, potentially on embedded devices with limited computing power (v. templates which can be analyzed post-sorting) \cite{Lin 2025B, Wu 2023, van der Molen 2024}.

To assess the statistical significance of performance differences across algorithms, we conducted one-way ANOVA tests for each metric followed by Tukey’s HSD post-hoc comparisons. For spike accuracy, both GC and MT significantly outperformed COM (p \textless 0.01). For template and spike runtime, MT was significantly slower than both COM and GC (p \textless 0.01), confirming its computational cost. For template and spike RMSE, GC showed significantly lower localization error than COM (p \textless 0.05), while differences between MT and GC were not statistically significant. No significant differences were found in template accuracy across algorithms (one-way ANOVA p = 0.0409; all pairwise comparisons p \textgreater 0.05).

\newgeometry{margin=1cm}
\begin{figure}[ht!]
    \centering
    \includegraphics[width=\dimexpr\linewidth\relax, clip]{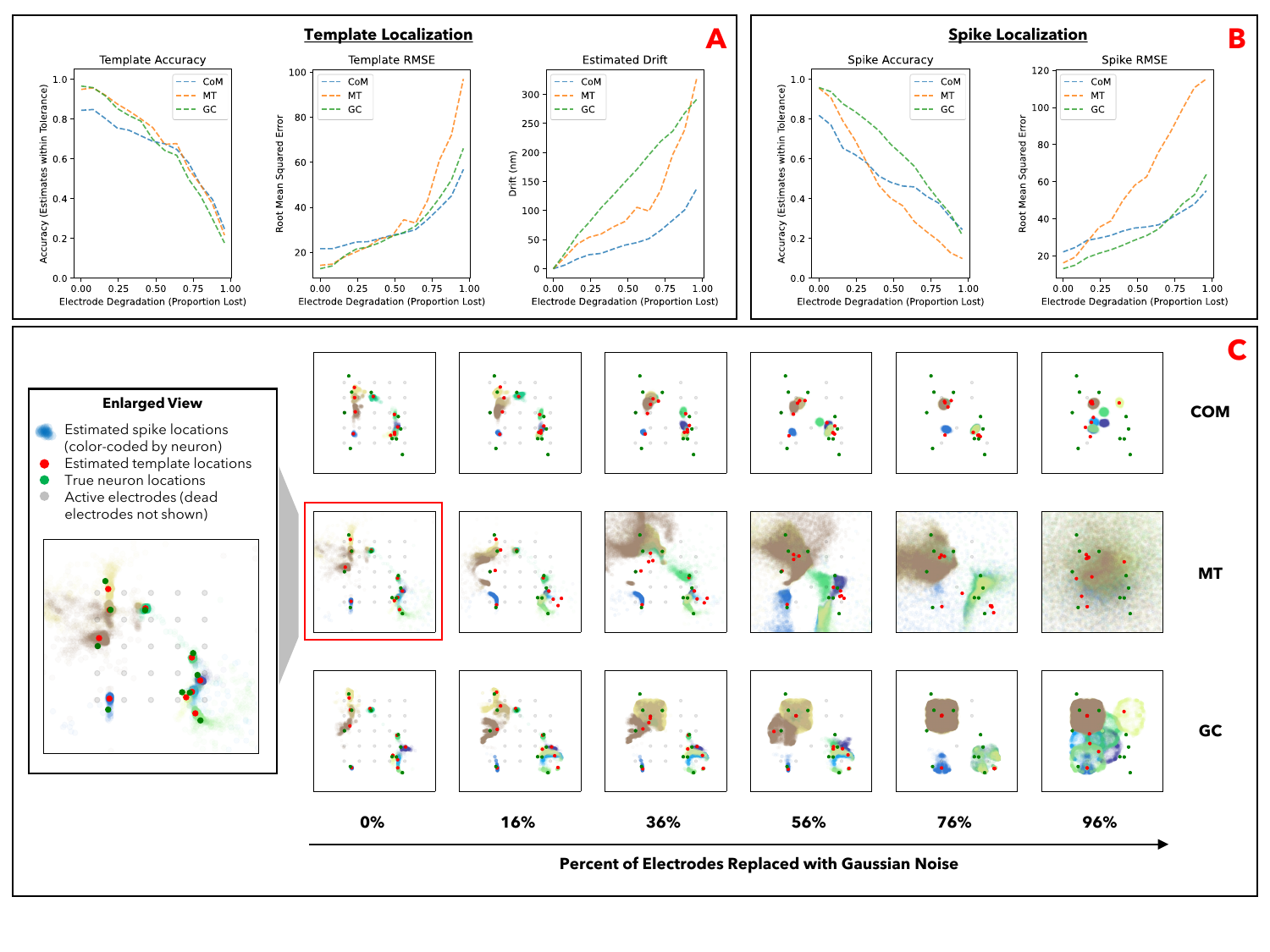}
    \caption{SSL algorithm robustness on simulated data (MEArec). \textbf{A}: Template localization performance metrics at varying levels of electrode degradation (measured as the percentage of electrodes replaced with Gaussian noise). Performance is assessed as (i) percentage of estimates within 30µm of actual locations (accuracy), (ii) Euclidean distance between estimates and actual location (RMSE), and (iii) estimated drift from original location (no drift occurs in ground truth data). \textbf{B}: Spike localization performance metrics at varying levels of electrode degradation. Performance is assessed via accuracy and RMSE. C: Visualization of template and spike localization estimates at varying levels of electrode degradation. Left panel is zoomed-in view of visualization outlined in red.}
    \label{fig:fig3}
\end{figure}
\restoregeometry

\newgeometry{margin=1cm}
\begin{figure}[ht!]
    \centering
    \includegraphics[width=\dimexpr\linewidth\relax, clip]{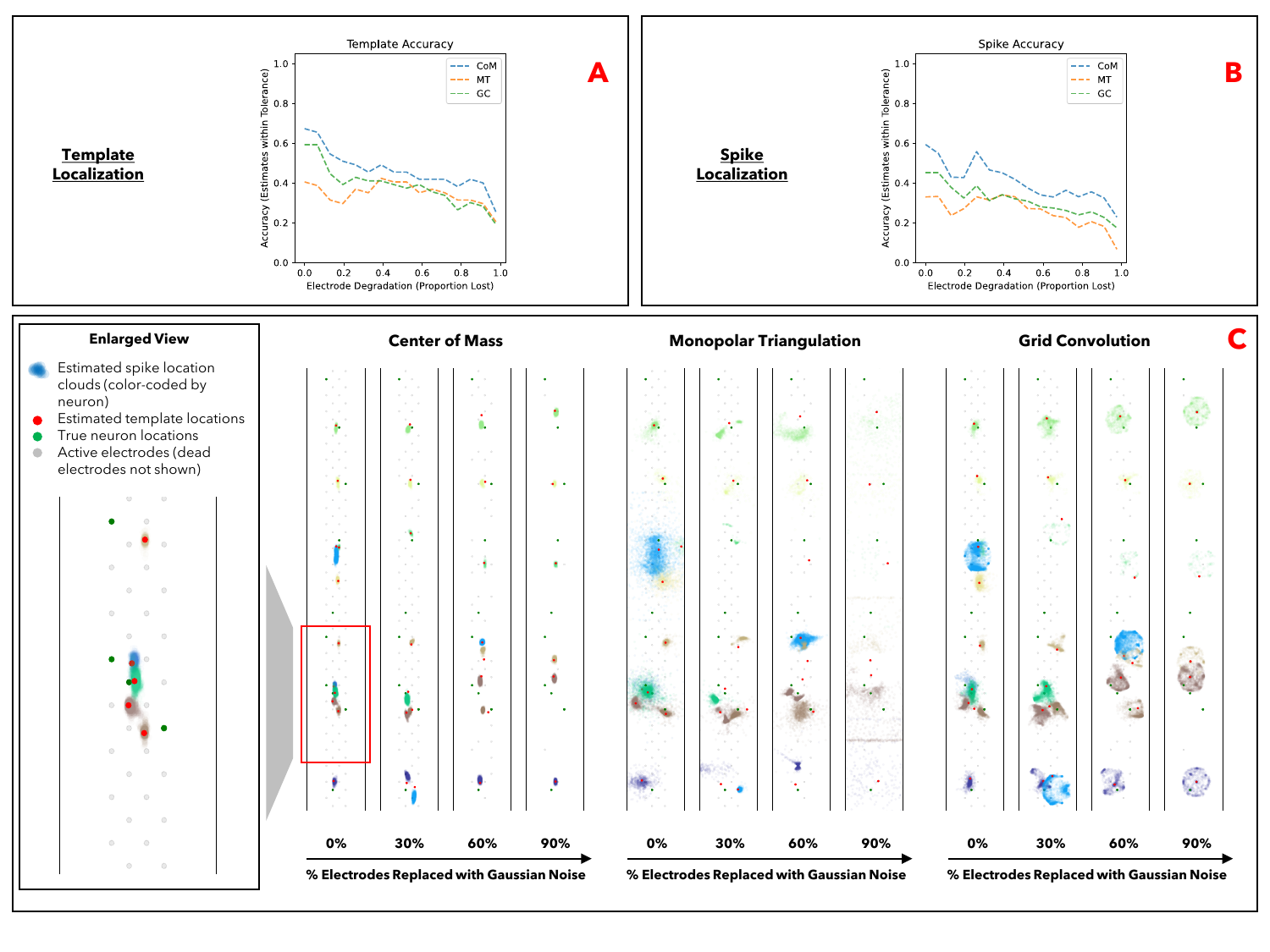}
    \caption{SSL algorithm robustness on experimental data (SPE-1). \textbf{A}: Template localization performance metrics at varying levels of electrode degradation (measured as the percentage of electrodes replaced with Gaussian noise). Performance is assessed as percentage of estimates within 80µm of actual locations. Since we have lower resolution actual location in experimental data v. simulated data, we use a higher tolerance for accuracy (80µm v. 30µm) and do not calculate the localization RMSE. \textbf{B}: Spike localization performance metrics at varying levels of electrode degradation. Performance is assessed via accuracy. \textbf{C}: Visualization of template and spike localization estimates at varying levels of electrode degradation. Left panel is zoomed-in view of visualization outlined in red.}
    \label{fig:fig4}
\end{figure}
\restoregeometry

\subsection*{Benchmarking results against decay}

We next benchmark the localization algorithms in the case of signal loss; i.e., how robust the localization algorithms are to electrode decay. In long-term neural recordings, individual electrodes will “fail” before the broader array, and over time the array will lose signal at an increasing proportion of its electrodes. In both the simulated ground truth and experimental ground truth datasets, we replace the signals of an increasing proportion of electrodes with Gaussian noise (the “dead electrodes”).

The deterioration of localization estimates as a result of electrode decay is visualized for the simulated ground truth dataset (Fig. 3C) and experimental ground truth dataset (Fig. 4C). COM is the most conservative, in that signal loss narrows the convex hull of the remaining electrodes, and COM estimates increasingly congregate towards the center of the array. GC is also conservative in that the estimates congregate in the grid around the remaining electrodes; this follows from the algorithm relying on the convolution between actual and theoretical signals at each electrode, and the convolution will be less meaningful around dead electrodes (i.e., will devolve to trivial convolution with Gaussian noise); intuitively, the theoretical and actual signals will overwhelmingly match around live electrodes v. Gaussian noise. MT is the least conservative; the algorithm solves the optimization problem with progressively less information, resulting in overfitting and a larger number of highly erroneous estimates (particularly for individual spikes). Our benchmarking on the simulated ground truth dataset (Fig. 3A, B) and experimental ground truth dataset (Fig. 4A, B) similarly show that COM and GC are more robust against electrode decay than MT.

Our robustness analysis additionally includes a measure of estimated drift from the original location (Fig. 3A). Since we are using the same ground truth data when simulating different degrees of electrode degradation, the neurons are located in the same positions relative to the array; this metric is a measure of stability, which is important in long-term recordings where we desire to consistently estimate neuron locations over time as electrodes decay (erroneous drift can result in the same neuron being identified as different neurons over long-term recordings).

\section*{Discussion}

This study provides a first-of-its-kind ground truth benchmarking of three widely used spike source localization algorithms: center of mass, monopolar triangulation, and grid convolution. As the field of brain-machine interfaces experience rapid growth and newer localization algorithms have been proposed with superior theoretical performance, it is important to ground these assessments in biological reality. Our study provides results for three algorithms, but also a ground-truth framework for benchmarking future localization algorithms.

\subsection*{Key Results \& Implications}

The results of our benchmarking indicate significant differences in the performance of the three localization algorithms. In the absence of electrode decay, GC and MT outperformed COM in the simulated ground truth dataset. This aligns with our expectations, as both GC and MT incorporate more physically realistic models compared to COM \cite{Hirabayashi 2004, Gold 2006}. However, we see MT performance suffer in the experimental ground truth dataset, highlighting the importance of real-world biological conditions in assessing a localization algorithm. When accounting for the electrode decay that occurs in long-term recording conditions, our findings demonstrate that COM and GC exhibit superior resilience compared to MT. As electrode failure rates increased, MT produced overfitting and larger localization errors, particularly for individual spike events. In contrast, COM’s simplicity and GC’s grid-based guardrails provided more stable estimates, making them more suitable for long-term recordings where electrode degradation is inevitable.

The superior robustness of COM and GC suggest better utility in long-term neural recordings, where resilience to electrode degradation and consistent spike localization are crucial for accurate longitudinal analysis. Localization results using COM and GC are likely to be more informative in spike sorting (as well as faster to calculate), and should better discriminate consistent neuron populations over time. In addition, the more reliable performance results are a more appropriate metric for assessing probe drift.

\subsection*{Future Direction}

We hope that the code for our benchmarking methodology, which will be publicly released and references publicly available datasets, will be informative for assessment of continued improvements in spike source localization algorithms. 

Moreover, we believe that spike source localization is a highly nascent field, where the use-cases are core to the advancement of brain-machine interfaces, but current algorithms are either based on simple heuristics or not sufficiently robust to biological noise. We note that at its core, the spike source localization problem is not new; it is a “triangulation” problem, which has been widely explored in fields such as geospatial localization techniques in telecommunications. Methods from geospatial applications, such as fuzzy logic and time-of-difference arrival techniques, have demonstrated superior performance in determining locations of wireless signal sources in real-world settings \cite{Teuber 2006, Zhou 2005, Liu 2007, Miao 2020, Zhang 2020}. By integrating principles from these fields, we believe there is significant opportunity to substantially improve the performance of spike source localization algorithms in brain-machine interfaces.

\section*{Supporting information}
All analysis scripts and supporting functions used in this study are available at https://github.com/haozhao1996/Spike-Localization-Algorithms. The code is released under the MIT License, permitting reuse, modification, and distribution with attribution. The simulated dataset is generated using MEArec (included in repository), and the experimental dataset is SPE-1 (publicly available).

\section*{Acknowledgments}
A.ML. acknowledges support from the RCC Fellowship of Harvard University and the Excellence Fellowship of the Fundacion Rafael del Pino. J.L. acknowledges support from the NSF EFRI. All authors would like to thank the Liu Group for helpful discussions. Computations were performed using resources provided by Harvard University.

\end{document}